\documentstyle[11pt,aas2pp4]{article}
\journalid{337}{15 January 1989}
\articleid{11}{14}

\slugcomment{Accepted, the Astrophysical Journal Letters - April/1997}

\begin{document}

\title{Discovery of optical pulsations in
V2116 Ophiuchi$\equiv$GX\,1+4\footnote{Based
on data collected at CNPq/Laborat\'orio Nacional de Astrof\'{\i}sica, Brazil}}

\author{Francisco J. Jablonski, Marildo G. Pereira, Jo\~ao Braga}
\affil{Divis\~ao de Astrof\'{\i}sica \\
       Instituto Nacional de Pesquisas Espaciais \\
       12201-970 S\~ao Jos\'e dos Campos, SP, Brazil \\
       chico@das.inpe.br, marildo@das.inpe.br, braga@das.inpe.br }

\author{Clemens D. Gneiding\altaffilmark{2}}
\affil{CNPq/Laborat\'orio Nacional de Astrof\'{\i}sica \\
       37500 Itajub\'a, MG, Brazil \\
       clemens@lna.br }

% There is a separate \altaffiltext for each alternate affiliation
% indicated above.
\altaffiltext{2}{Also Instituto Nacional de Pesquisas Espaciais, Brazil }

\begin{abstract}

We report the detection of pulsations with $\sim 124$ s period in
V2116 Oph, the optical counterpart of the low-mass X-ray binary
GX\,1+4. The pulsations are sinusoidal with modulation amplitude of up
to 4\% in blue light and were observed in ten different observing
sessions during 1996 April-August using a CCD photometer at the 1.6-m
and 0.6-m telescopes of Laborat\'orio Nacional de Astrof\'{\i}sica, in
Brazil. The pulsations were also observed with the $UBVRI$ fast
photometer.  With only one exception the observed optical periods are
consistent with those observed by the BATSE instrument on board the
{\sl Compton Gamma Ray Observatory} at the same epoch.  There is a
definite correlation between the observability of pulsations and the
optical brightness of the system: V2116~Oph had $R$ magnitude in the
range $15.3-15.5$ when the pulsed signal was detected, and $R =
16.0-17.7$ when no pulsations were present.  The discovery makes
GX\,1+4 only the third of $\sim 35$ accretion-powered X-ray pulsars to
be firmly detected as a pulsating source in the optical.  The presence
of flickering and pulsations in V2116 Oph adds strong evidence for an
accretion disk scenario in this system. The absolute magnitude of the
pulsed component on 1996 May 27 is estimated to be $M_V \sim
-1.5$. The implied dimensions for the emitting region are $1.1
R_{\sun}$, $3.2 R_{\sun}$, and $7.0 R_{\sun}$, for black-body spectral
distributions with $T = 10^5$ K, $2 \times 10^4$ K, and $1 \times
10^4$ K, respectively.

\end{abstract}

\keywords{stars: individual (V2116 Oph) --- pulsars: individual (GX\,1+4)
--- stars: oscillations --- stars: neutron --- X-rays: stars}

\section{Introduction}
GX\,1+4 is a low-mass X-ray binary (LMXB) well known for its unique
characteristics such as extended high/low states (McClintock \&
Leventhal 1989; Mony et al. 1991; Makishima et al. 1988), luminosity
of up to $\sim 10^{38}$ erg/s in the high state (Rao et al. 1994;
Nagase 1989), the hardest spectrum of all X-ray binaries (with kT
$\sim35-55$ keV in a thermal bremsstrahlung model) (Laurent et
al. 1989; Staubert et al. 1995; Predehl, Friedrich \& Staubert 1995)
and the fastest recorded rate of change of period amongst the
persistent X-ray pulsars ($\dot{P}/P\sim -2\%$/yr in the 1970s)
(Nagase 1989). The optical counterpart of GX\,1+4 was identified by
Glass \& Feast (1973) and confirmed recently by an improved X-ray
position obtained with the ROSAT satellite (Predehl et al. 1995).
Davidsen et al. (1977) showed that V2116 Oph has many characteristics
of a symbiotic system, with the late-type star being classified as M6
III. Shahbaz et al. (1996) classify the late-type spectrum as M5
III. The absorption spectrum of V2116 Oph redward of H$\alpha$ is very
similar to the spectrum of NSV 11776 (Cieslinski, Elizalde \& Steiner
1994) which is classified as M4 III or later. Davidsen et al. (1977)
point out a well defined excess of blue light in the spectrum, and
that H$\alpha$, as well as the (blue) continuum, show variability.

Recent low-time-resolution photometric studies have shown interesting
results like the approximately simultaneous H$\alpha$ and X-ray
flaring activity (Manchanda et al. 1995; Greenhill et al. 1995) and
the very low levels of H$\alpha$ emission (Sood et al. 1996) following
the X-ray low-state in which GX\,1+4 entered on 1996 August 18
(Chakrabarty \& Prince 1996).

Despite all of the interesting features shown by the X-ray source,
surprisingly little optical photometry work has been done with good
time resolution. In the sole short-term variability study of this
system performed so far (Krzeminski \& Priedhorski 1978), a search for
optical pulsations in H$\alpha$ established upper limits of 1.7\% and
0.7\% to the pulsed fraction of sinusoidal and X-ray pulse shapes in
the range 111 s $ < P < $ 143 s. This null result was interpreted as
due to smearing and suppression effects due to the large dimension of
the H$\alpha$ emitting region, $r_{H\alpha}\sim 6\times 10^{13}$ cm,
as calculated by Davidsen et al. (1977). The escape time for H$\alpha$
photons with an optical depth $\tau_{H\alpha} \sim 10^2$ from a cloud
of that size would be $(r_{H\alpha}/c) \tau_{H\alpha} \sim 2 \times
10^5$ sec.

Motivated by the discovery of flickering with time scales of
$\sim$minutes (Braga et al. 1993), and night-to-night variations on
the mean $R$ magnitude of up to 0.5 mag, we are conducting a
concentrated observational effort trying to answer fundamental
questions about this system: what is its orbital period? is this a
wind-fed or accretion-disk-fed pulsar? What is the origin of the
photometric variability in long and short time scales?

\section {Observations and Data Reduction}

The data discussed in this paper were obtained with the 1.6-m and
0.6-m telescopes of Laborat\'orio Nacional de Astrof\'{\i}sica, in
Brazil. The CCD photometer we have built consists of a Wright
Instr. thermoelectrically-cooled camera with a EEV CCD-02-06
back-illuminated chip operated in the frame-transfer mode (385
$\times$ 289 pixels in the image section).  The timing for the
instrument is provided by a Datum Inc. BC627AT Global Position System
(GPS) receiver and timing board installed in the same IBM-PC clone
used to read the CCD camera.  Integration times down to 1 second are
possible, especially with $2 \times 2$ on-chip binning.

The plate scale produced by the various combinations of telescopes,
focal reducers and binning factors is always
$\lesssim1\farcs2$/pixel. The field around V2116 Oph is relatively
uncrowded, permitting safe aperture photometry to monitor its
brightness relative to several comparison stars.

The reduction of the data is presently done off-line using a set of
IRAF\footnote{IRAF is distributed by National Optical Astronomy
Observatories, which is operated by the Association of Universities
for Research in Astronomy, Inc., under contract with the National
Science Foundation.} scripts written by one of us (FJ). The reduction
of a few thousand images, including the standard preparation
procedures (bias, dark and flat-field) can be done in the day
following the observations.

The data summarized in Table 1 resulted from 20,465 images in the
field of V2116 Oph.  The CuSO$_4$ filter used is a 4mm-thickness
liquid solution similar to what we have used for many years in the $U$
passband of our $UBVRI$ photometer (Jablonski et al. 1994). This
non-standard passband maximizes photon count without sacrificing too
much the capability of doing differential photometry.  Column 3 of
Table 1 shows the integration time $\Delta$t, and column 4 shows the
total duration of each run. On 1996 April 26 for instance, the data
set consists of 3300 sequential images of 5 s integration time each.

Each night, a few images in the $R$ filter were obtained to provide a
long-term light curve of the system. Since we have determined
photoelectrically the magnitude of star \#9 in the chart of Doxsey et
al. (1977) ($V=13.39\pm0.04$, $V-R=+0.77\pm0.02$, $R-I=0.79\pm0.01$)
even nights of poorer photometric quality provide good magnitude
measurements for V2116 Oph.  This is the same comparison star used in
the H$\alpha$ photometry of Greenhill et al. (1995). Column 5 of Table
1 lists the mean $R$ magnitude of V2116 Oph in each observational run.

\section{Analysis}
Figure 1 shows the light curve of V2116 Oph obtained with the CuSO$_4$
filter on 1996 April 26, where the optical pulsations with 124 s were
first detected (Jablonski et al. 1996). The comparison stars observed
simultaneously (two of them are shown in this figure) provide a very
good estimate of the noise in the variable star's light curve.

Figure 2 shows the power spectrum of the light curve in Figure 1.  The
dotted line marks the expected level of noise estimated from the light
curves of six comparison stars measured simultaneously.  The isolated
feature close to 0.008 Hz has peak power $\sim200$ times the power in
the local continuum. A discrete periodogram with increased frequency
resolution gives a better estimate of the period of that feature:
124.17$\pm$0.04 s. There are two interesting additional pieces of
information in Figure 2: the inclination of the power spectrum at low
frequencies, $\alpha =2.01\pm0.05$ in ${\rm Power} \propto
f^{-\alpha}$, and the period where this component intercepts the noise
level, $P_0=57\pm2$ s.  The flickering behavior of V2116\,Oph is very
similar to that shown by cataclysmic variables (Bruch 1992). An
estimate of the amount of flickering in the light curves is given in
the last column of Table 1. It measures the excess power above the
photon noise level, integrated in the frequency range $0.001-0.005$
Hz. We follow van der Klis (1989) and express this quantity in terms
of percentage r.m.s. variation in the light curve.

The detection of stable optical pulsations with period close to what
is expected from the X-rays history of GX\,1+4 prompted us to observe
the system as often as possible. Table 1 shows a summary of all
subsequent observations. In the case of a detection, the barycentric
period and the semi-amplitude of the pulsed signal are shown in
columns 6 and 7.  In Figure 3 we plot the periods listed in Table 1
together with the results from the continuous monitoring by BATSE. The
very good quality of the X-rays daily period determinations results
from the large modulation fraction and from the use of five-day
segments in the analysis (Chakrabarty et al. 1997).

\section{Discussion}

\noindent We can summarize the results of this paper as follows:

We have detected optical pulsations in V2116 Oph with periods very
close to the expected from the X-rays history of GX\,1+4 (Chakrabarty
et al. 1997).  This adds a third object to the small group of
accreting pulsars that are LMXB {\it and} present optical
pulsations. The other cases are HZ Her$\equiv$Her-X1 (Davidsen et al.
1972; Middleditch \& Nelson 1976) and KZ TrA$\equiv$4U1626-67
(Ilovaisky, Motch \& Chevalier 1978; Middleditch et al. 1981).

We interpret the presence of flickering in the light curves as
evidence of an accretion disk in the system. The power spectrum shown
in Figure 2 indicates that the variations due to flickering appear on
time scales as short as 57 s before they get immersed on the flat
noise component due to photon statistics. If we associate the shortest
time scale for flickering in our data, namely 57 s, with the radius of
a Keplerian orbit in the accretion disk, we obtain a dimension $r
\sim2.5 \times 10^9$ cm, for a $1.4 M_{\sun}$ neutron star. In other
words, if we associate $r$ with an upper limit on the magnetospheric
radius $r_M$ (Frank, King \& Raine 1992), then the fastness parameter
$\omega_s$ in the theory of Ghosh \& Lamb (1979a,b) would be $\approx
0.5$. This is consistent with $\omega_s < 1$ needed for steady
accretion. The standard theory (Frank et al. 1992) also gives an
estimate of the surface magnetic field of the neutron star, $B_0
\approx$ 10$^{14}$ G, assuming a typical luminosity of 10$^{37}$ erg
s$^{-1}$. This is consistent with a previous estimate by White (1988).
 
The absence of flickering when the system is faint (the total
r.m.s. variation above photon noise in the light curve of June 19 is
$\lesssim 1\%$) indicates that the accretion may stop
episodically. This is consistent with the observed low states in
X-rays (Chakrabarty \& Prince 1996) contemporaneous with very low
levels of H$\alpha$ emission (Sood et al. 1996).  If Roche lobe
overflow is the preferred accretion mode in GX1+4, we are forced to
conclude that the M companion does not fill its lobe all the
time. This is expected if the M-giant presents large-scale mass
motions it its outer layers, like red irregular and semi-regular
variables usually do (Querci 1986). Besides providing episodes of low
accretion, semi-regular variations in the accretion flux could also be
a good explanation for the so called ``300 days'' cycle in the torque
history of the neutron star (Cutler, Dennis \& Dolan 1986; Chakrabarty
et al.  1997).

The origin of the optical pulsed emission cannot be determined from
the data discussed in this paper, but taking what happens in HZ Her
and KZ TrA as a paradigm, it could be a combination of reprocessing of
X-rays in the accretion disk, in the atmosphere of late-type star, and
in the stream between the latter and the disk (Middleditch \& Nelson
1976; Middleditch et al. 1981). From the observed colors of the DC and
pulsed light on 1996 May 27, we estimate the absolute magnitude of the
pulsed component to be $M_V \sim -1.5$ for a distance of $\sim$ 10 kpc
and $E(B-V)$ in the range $1.8-2.0$ (Jablonski \& Pereira 1997).  If
the pulsed light follows a black-body spectral distribution with
temperature $T_{BB} = 10^5$ K, the size of the emitting region would
be $\sim 1.1R_{\sun}$. For lower temperatures, $2 \times 10^4$ K and
$1 \times 10^4$ K, we obtain dimensions of $3.2 R_{\sun}$ and $7.0
R_{\sun}$, respectively. The correspondent areas are not prohibitive
either as fractions of the surface area of the companion star or as
fractions of an accretion disk area. The accretion stream is ruled out
as a likely sole source of the pulsations because its projected area
is too small.  If the pulsations show a dependence with orbital phase,
like in HZ Her and KZ TrA, the long term monitoring of the optical
pulsations may be useful to answer a long-lasting (and very basic)
question about this system: what is the orbital period of GX\,1+4 ? So
far, we only know that in order to fill the corresponding Roche lobe,
a M4-6 III star with 80-100 $R_{\sun}$ would need $P_{\rm orb} \gtrsim
100$ days in a binary with mass ratio $q \sim 1$.

The BATSE measurements (Chakrabarty et al. 1997) show that it is very
difficult to measure the orbital period of the system from the Doppler
effect on the X-ray pulses alone, by the following reasons: the
orbital period is long, probably $\gtrsim 100$ d, the orbital
inclination may be low (though this is less certain), and it is very
difficult to separate the effects of the spin-down (and its
fluctuations) from the Doppler effect itself on a long-period
orbit. Our best determination of pulse period, namely 124.17$\pm0.04$
s, on April 26, is significantly different from the BATSE measurement,
123.9453 $\pm$0.0014 s, based on a linear interpolation between the
daily values of Chakrabarty et al. (1997). If caused by Doppler
effect, the implied difference in velocity between the X-ray and
optical emitting regions would be $\sim540(\sin)^{-1} i$ km/s, too
high for a binary with the characteristics discussed earlier.
 
By analogy with HZ\,Her and KZ\,TrA we expect to see sidebands of
$P_{\rm opt}$ or $P_{\rm X}$ according to $P_{\rm opt}^{-1} = P_{\rm
X}^{-1} - P_{\rm orb}^{-1}$.  If we try to explain the observed
difference between $P_{\rm opt}$ and $P_{\rm X}$ in terms of
sidebands, then the derived $P_{\rm orb}$ is too small, $\sim 0.8$
days, again inconsistent with our previous knowledge about this
system. A good observational coverage will settle this point, in
particular if we are able to track the {\em phase} of the pulsations
from simultaneous measurements in the optical and X-ray bands.

The relatively long period and large amplitude of the optical
pulsations in V2116 Oph open interesting perspectives for future
work. Likely projects include the detailed history of the pulsations
on a night-to-night basis and its correlation with the X-ray timings,
a spectrophotometric study of the optical and NIR spectrum of the
pulsations and the correlation of the visibility of the pulsations
with orbital phase.

\acknowledgements 
MP acknowledges the Brazilian agencies CNPq and CAPES for graduate
studies support; FJ and JB acknowledge CNPq for research fellowships.
We thank Dr. Deepto Chakrabarty for calling our attention to the
incorrect error bar for the 124\,s period in Jablonski et al. (1996).
The correct value is shown in Table 1 of this paper.

\clearpage

\figcaption{Differential photometry CCD light curve of V2116 Oph on
1996 April 26, with 5 s time resolution. The differential light curves
of two nearby stars are shown for comparison. The brightest star is
object \# 10 in the chart of Doxsey et. al (1977).  The inset shows a
section where individual pulses can be seen. The arrows pointing to
subsequent maxima were produced by constant-period (124.17 s)
prediction. The small bar in the upper right corner of the inset shows
the estimated level of noise in that section of the light curve.}

\figcaption{The power spectrum of the light curve in Figure 1. The
horizontal dashed line shows the estimated level of noise derived from
the light curves of six comparison stars measured simultaneously. The
inset shows the light curve folded on the 124.17 s period. The phase
interval is repeated twice for better visualization and the vertical
scale is relative amplitude.}

\figcaption{The period of the optical pulsations listed in Table 1
(filled circles) together with the BATSE daily measurements (crosses;
Chakrabarty et al. 1997).}

\clearpage

\begin{deluxetable}{lccccccc}
%\tablewidth{33pc}
\tablewidth{0pc}
\tablecaption{Fast Photometry of V2116 Oph}
\tablehead{
\colhead{Date}   & \colhead{Filter} & \colhead{$\Delta$t} &
\colhead{Duration} & \colhead{R} & \colhead{Barycentric period}   &
\colhead{Semi-amplitude} & \colhead{Flickering\tablenotemark{f}} \\ 
\colhead{(1996)} & \colhead{}       & \colhead{(s)}     &
\colhead{(hrs)}    & \colhead{(mag)}  & \colhead{(s)} &
\colhead{(\%)} & \colhead{(\%)}            } 

\startdata
 Apr 26\tablenotemark{a} & CuSO$_4$ & 5 & 4.6 & 15.48 &
 ~124.17$\pm$0.04 & 1.3$\pm$0.1 & 2.3 \nl May 21\tablenotemark{b} &
 CuSO$_4$ & 15 & 4.2 & 15.26 & 124.1$\pm$0.3 & 0.6$\pm$0.3 & 2.5 \nl
 May 25\tablenotemark{c} & CuSO$_4$ & 20 & 0.4 & 15.29 & 124.6$\pm$1.6
 & 1.8$\pm$0.5 & 0.7 \nl May 26\tablenotemark{c} & CuSO$_4$ & 10 & 0.9
 & 15.29 & 124.2$\pm$0.9 & 1.2$\pm$0.4 & 1.9 \nl May
 27\tablenotemark{c} & CuSO$_4$ & 20 & 0.9 & 15.28 & 124.0$\pm$0.3 &
 4.3$\pm$0.5 & 1.7 \nl May 27\tablenotemark{c} & R & 20 & 0.8 & 15.28
 & 124.1$\pm$0.4 & 2.6$\pm$0.4 & 2.0 \nl May 27\tablenotemark{ad} &
 Clear & 15\tablenotemark{e} & 0.7 & 15.28 & 124.5$\pm$0.5 &
 2.5$\pm$0.4 & 1.4 \nl May 27\tablenotemark{ad} & U &
 15\tablenotemark{e} & 0.7 & 15.28 & \nodata & $<$12 & 8.6 \nl May
 27\tablenotemark{ad} & B & 15\tablenotemark{e} & 0.7 & 15.28 &
 \nodata & $<$4.4 & 4.8 \nl May 27\tablenotemark{ad} & V &
 15\tablenotemark{e} & 0.7 & 15.28 & 124.1$\pm$1.0 & 5.3$\pm$1.6 & 4.6
 \nl May 27\tablenotemark{ad} & R & 15\tablenotemark{e} & 0.7 & 15.28
 & 124.6$\pm$0.6 & 3.3$\pm$0.6 & 1.7 \nl May 27\tablenotemark{ad} & I
 & 15\tablenotemark{e} & 0.7 & 15.28 & 124.1$\pm$0.6 & 2.2$\pm$0.4 &
 1.3 \nl May 28\tablenotemark{c} & CuSO$_4$ & 20 & 1.5 & 15.37 &
 123.9$\pm$0.5 & 1.7$\pm$0.6 & 2.5 \nl May 28\tablenotemark{c} & R &
 20 & 1.7 & 15.37 & 124.3$\pm$0.5 & 0.6$\pm$0.2 & 1.1 \nl Jun
 19\tablenotemark{a} & Clear & 5 & 7.0 & 17.72 & \nodata &$<$0.05 &
 0.1 \nl Jun 20\tablenotemark{a} & Clear & 5 & 6.6 & 17.68 & \nodata
 &$<$0.07 & 0.2 \nl Jul 13\tablenotemark{b} & R & 25 & 3.7 & 17.01 &
 124.1$\pm$0.2 & 0.9$\pm$0.3 & 1.2 \nl Aug 04\tablenotemark{b} & R &
 20 & 3.0 & 15.67 & 124.2$\pm$0.3 & 0.4$\pm$0.2 & 0.5 \nl Aug
 05\tablenotemark{b} & R & 20 & 4.4 & 15.45 & 124.0$\pm$0.4 &
 0.2$\pm$0.2 & 1.3 \nl Aug 06\tablenotemark{b} & CuSO$_4$ & 20 & 3.3 &
 15.28 & 124.7$\pm$0.3 & 0.4$\pm$0.1 & 0.6 \nl Aug 07\tablenotemark{b}
 & R & 20 & 1.7 & 15.29 & \nodata &$<$0.1 & 0.5 \nl Aug
 18\tablenotemark{b} & Clear & 10 & 1.4 & 15.99 & \nodata &$<$0.15 &
 0.2 \nl Aug 19\tablenotemark{b} & R & 20 & 3.0 & 16.29 & \nodata
 &$<$0.33 & 0.5 \nl Aug 20\tablenotemark{b} & R & 25 & 4.2 & 16.09 &
 \nodata &$<$0.17 & 0.3 \nl Aug 22\tablenotemark{b} & R & 25 & 5.7 &
 16.94 & \nodata &$<$0.16 & 0.9 \nl
\enddata

\tablenotetext{a}{1.6-m telescope}
\tablenotetext{b}{0.6-m Zeiss telescope}
\tablenotetext{c}{0.6-m Boller \& Chivens telescope}
\tablenotetext{d}{FOTRAP photometer}
\tablenotetext{e}{Time resolution instead of integration time}
\tablenotetext{f}{Integrated in the range $0.001-0.005$ Hz}

\end{deluxetable}


\clearpage

\begin{references}
% \mnras, \aap, \apj, \pasp, \apjl, \apjs, \aj
%
\reference {} Braga, J., Jablonski, F., D'Amico, F., Elizalde, F., \& Steiner, 
           J. 1993, {Rev. Mex. Astr. Astrof.}, 26, 113
\reference {} Bruch, A. 1992, \aap, 266, 237
\reference {} Chakrabarty, D. \& Prince, T. 1996, IAU Circ. No. 6478
\reference {} Chakrabarty, D., Bildsten, L., Grunsfeld, J. M., Koh, D. T.,
           Nelson, R. W., Prince, T. A., Vaughan, B. A., Finger, M. H., 
           \& Wilson, R. B. 1997, \apjl, submitted
\reference {} Cieslinski, D., Elizalde, F., \& Steiner, J. E. 1994, \aaps, 106, 243
\reference {} Cutler, E. P., Dennis, B. R., \& Dolan, J. F. 1986, \apj, 300, 551
\reference {} Davidsen, A., Henry, J. P., Middleditch, J., \& Smith, H. E. 1972,
           \apjl, 177, L97
\reference {} Davidsen, A., Malina, R., \& Bowyer, S. 1977, \apj, 211, 866
\reference {} Doxsey, R.E., Apparao, K.M.V., Bradt, H.V., Dower, R.G., 
           \& Jernigan, J.G. 1977, \nat, 270, 586
\reference {} Frank, J., King, A., Raine, D. 1992, Accretion Power in
           Astrophysics, 2nd ed., (Cambridge: Cambridge Univ. Press)
\reference {} Ghosh, P. \& Lamb, F. K. 1979a, \apj, 232, 259
\reference {} Ghosh, P. \& Lamb, F. K. 1979b, \apj, 234, 296
\reference {} Glass, I.S., \& Feast, M. W. 1973, { Nature Phys. Sci.}, 245, 39
\reference {} Greenhill, J.G., Watson, R.D., Tobin, W., Pritchard, J.D., \&
           Clark, M. 1995, \mnras, 274, L61
\reference {} Ilovaisky, S. A., Motch, C., \& Chevalier, C. 1978, \aap, 70, L19
\reference {} Jablonski, F., Baptista, R., Barroso Jr., J., Gneiding, C. D.,
           Rodrigues, F., \& Campos, R. P. 1994, \pasp, 106, 1172
\reference {} Jablonski, F., Pereira, M. G., Braga, J., \& Gneiding, C. D. 1996,
           IAUC Circ. No. 6489
\reference {} Jablonski, F., \& Pereira, M. G. 1997, in preparation
\reference {} Querci, F. R. 1986, in The M-type Stars,
            ed. S. Jordan \& R. Thomas,  (NASA SP-492)(Washington: U.S. Government Printing Office), 1
\reference {} Krzeminski, W. \& Priedhorski, W.C. 1978, \pasp, 106, 566
\reference {} Laurent, P., Salotti, L., Paul, J., Lebrun, F., Denis, M.,
           Barret, D., Jourdain, E., Roques, J. P., Churazov, E., Gilfanov, M.,
           Sunyaev, R., Diachkov, A., Khavenson, N., Novikov, B., Chulkov, I., \&
           Kuznetzov, A. 1993, \aap, 278, 444
\reference {} Makishima, K., Ohashi, T., Sakao, T., Dotani, T., Inoue, H.,
           Koyama, K., Makino, F., Mitsuda, K., Nagase, F., Thomas, H. D., Turner,
           M. J. L., Kii, T., \& Tawara, Y. 1988, \nat, 333, 746
\reference {} Manchanda, R.K., James, S.D., Lawson, W.A., Sood, R.K.,
           Grey, D.J., \& Sharma, D.P. 1995, \aap, 293, L29
\reference {} McClintock, J.\ E., Leventhal, M. 1989, \apj, 346, 143
\reference {} Middleditch, J., \& Nelson, J. E. 1976, \apj, 208, 567
\reference {} Middleditch, J., Mason, K. O., Nelson, J. E., \& White, N. E. 1981,
           \apj, 244, 1001
\reference {} Mony, B., Kendziorra, E., Maisack, M., Stauber, R., Englhauser,
           J., D\"obereiner, S., Pietsch, W., Reppin, C., Tr\"umper, J., Churazov, E. M.,
           Gilfanov, M. R., \& Sunyaev, R. 1991, \aap, 247, 405
\reference {} Nagase, F. 1989, \pasj, 41, 1
\reference {} Predehl, P., Friedrich, S. \& Staubert, R. 1995, \aap, 294, L33
\reference {} Rao, A. R., Paul, B., Chitnis, V. R., Agrawal, P. C., and
           Manchanda, R. K. 1994, \aap, 289, L43
\reference {} Sood, R., James, S., Lawson, W., Manchanda, R., and
           Heisler, C. 1996,  IAU Circ. No. 6496
\reference {} Shahbaz, T., Smale, A. P., Naylor, T., Charles, P. A., van Paradijs, J.,
           Hassal, B. J. M., \& Callanan, P. 1996, \mnras, 282, 1437
\reference {} Staubert, R., Maisack, M., Kendziorra, E., Draxler, T., Finger,
           M. H., Fishman, G. J., Strickman, M. S., \& Starr, C. H. 1995, 
           Adv. Space Res. 15, (5)119
\reference {} van der Klis, M. 1989, in Timing Neutron Stars, ed. H. Ogelman \&
              E. P. J. van den Heuvel (NATO ASI Ser. C, 262)(Dordrecht: Kluwer), 203
\reference {} White, N. E. 1988, \nat, 333, 708
%
\end{references}
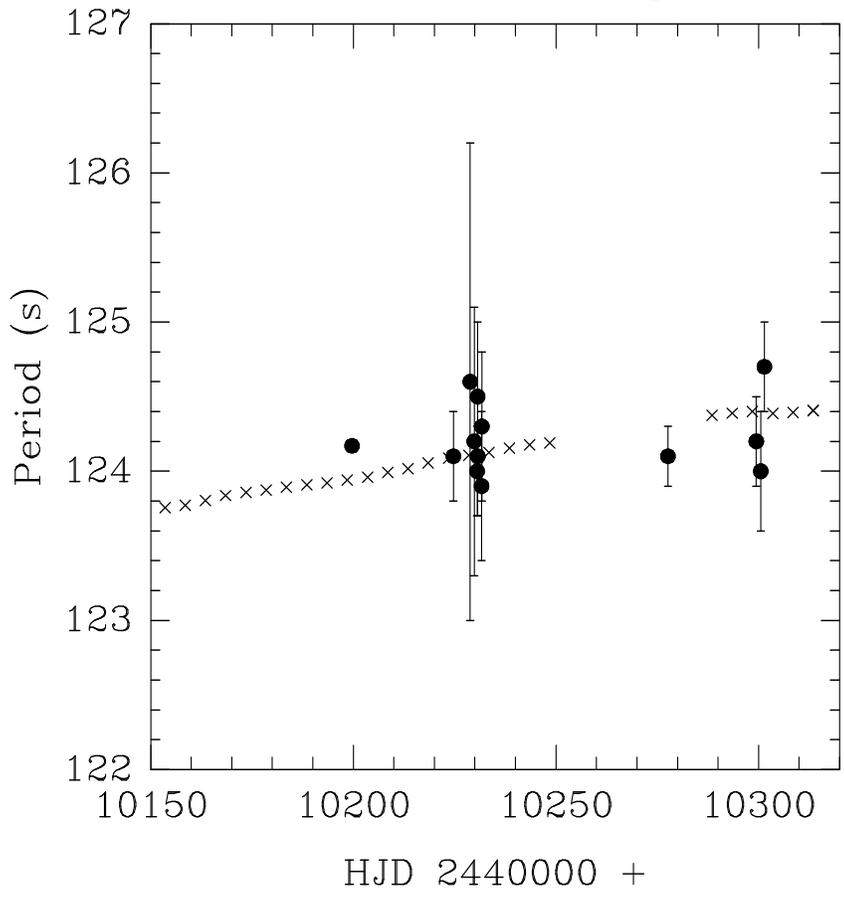
\end{document}